\documentclass[]{emulateapj}
\shorttitle{Imbalanced Turbulence}
\begin{document}
\title{Structure of Stationary Strong Imbalanced Turbulence}
\author{A. Beresnyak, A. Lazarian}
\affil{Dept. of Astronomy, Univ. of Wisconsin, Madison, WI 53706}
\email{andrey, lazarian@astro.wisc.edu}

\def\L{{\Lambda}}
\def\l{{\lambda}}

\begin{abstract} 
  In this paper we systematically study the spectrum and structure of
  incompressible MHD turbulence by means of high resolution direct
  numerical simulations.  We considered both balanced and imbalanced
  (cross-helical) cases and simulated sub-Alfv\'enic as well as
  trans-Alfv\'enic turbulence. This paper extends numerics preliminarily reported in
  Beresnyak \& Lazarian 2008. We confirm that driven imbalanced
  turbulence has a stationary state even for high degrees of
  imbalance. Our major finding is that the structure of the dominant
  and subdominant Alfv\'enic components are notably different.
  Using the most robust observed quantities, such as the energy
  ratio, we believe we can reject several existing models of
  strong imbalanced turbulence.
\end{abstract}

\keywords{MHD -- turbulence -- ISM: kinematics and dynamics}

\section{Introduction}

Astrophysical fluids are known to be turbulent from large scales
of the order of hundreds of kiloparsecs,
as in the case of galaxy clusters (Vogt \& Ensslin 2005), to scales
as small as hundreds of kilometers. Armstrong et al. (1995)
reported a wide range of electron density
fluctuations probed by scintillation and scattering
techniques. New statistical techniques using
Doppler-shifted lines have enabled studies of velocity fluctuations
on scales larger than a fraction of a parsec (see Lazarian (2009) and refs. therein).
Recent years have been marked by new understanding of the key role that
turbulence plays in a number of astrophysical processes (Cho,
Lazarian \& Vishniac 2003, Elmegreen \& Scalo 2004). Most notably,
turbulence has drastically changed the paradigms of interstellar
medium and molecular cloud evolution (Stone et al 1998, Ostriker et al. 2001,
Vasquez-Semadeni et al. 2007, see also review by McKee \& Ostriker
2007). Small scale turbulence has been probed by a variety of
approaches such as gyrokinetics, Hall MHD and electron MHD
(Howes et al 2006, Schekochihin et al 2007, Galtier et al 2003,
Cho \& Lazarian 2004). This progress calls for better understanding
of the fundamentals of turbulence. One reason for doing this is to understand to
what extent the turbulence simulated with diffusive computer codes
resembles actual turbulence in astrophysical fluids with low
diffusivity.

We start with a few general remarks on the relation of fluid dynamics
and description of astrophysical fluids. The continuous fluid
description has been successful describing a wide range of physical
phenomena. In space, due to the presence of ionizing radiation and
cosmic rays (CRs) the medium is almost always ionized to some degree,
i.e. is a plasma. Although the dynamics of plasma is complicated on
small scales, it can be considered well-conducting continuous fluid on
large scales and, therefore, Magnetohydrodynamics or MHD is
applicable.

One of the most interesting phenomena in fluids is turbulence, a
seemingly random stochastic flow which appears spontaneously as long
as the microscopic dissipation coefficients such as viscosity or
magnetic diffusivity are small (which correspond to large Reynolds or
magnetic Reynolds numbers). Turbulence increases
dissipation due to so-called turbulent cascade, a nonlinear transfer
of energy to smaller scales. As astrophysical fluids are turbulent,
this affects dynamics and is manifested in a variety of situations
such as reconnection, momentum transfer in accretion disks, etc.

The basic theoretical study of the complicated nonlinear dynamics of
turbulence has been concentrated on incompressible turbulence with
large Reynolds number (Kolmogorov, 1941).  While dissipation in plasma
might be much more complex than dissipation in molecular
gases\footnote{The dissipation in collisionless plasmas has been an
  area of active research recently, see, e.g. Howes et al 2008 and
  ref. therein.}, in the asymptotic large Reynolds number flows the
``inertial range'' fluctuations do not feel the peculiarities of the
dissipation. On the other hand, it is often necessary to account for
compressibility of the fluid, as it could be significant.
An example of which
is the interstellar medium which has has sonic Mach numbers in a range of
$0.1-10$. In this case the dynamics of Alfv\'en and slow-mode
perturbations on small scales can be considered incompressible (Lithwick
\& Goldreich 2001), while
fast mode has the dynamics of its own (Cho \& Lazarian 2002, 2003). On
the other hand, even at large scales where the compressibility is
significant, one can find the features of incompressible dynamics
(see, e.g., Beresnyak, Lazarian \& Cho 2005). In other words, the
understanding of incompressible MHD turbulence, which is often called
Alfv\'enic turbulence due to the dominant role the Alfv\'en perturbations
play in the cascade, is of utmost importance.

Ideal incompressible MHD equations can be written in the following
simple form
$$\partial_t{\bf w^\pm}+\hat S ({\bf w^\mp}\cdot\nabla){\bf w^\pm}=0, $$
where $\hat S$ is a solenoidal projection operator and Elsasser
variables are defined in terms of velocity $\bf{v}$
and magnetic field in velocity units ${\bf b=B}/(4\pi \rho)^{1/2}$
as ${\bf w^+=v+b}$ and ${\bf w^-=v-b}$.
Although fairly idealized, the problem of incompressible MHD 
turbulence with large Reynolds number has been difficult.
First attempts to treat it
was an IK model by Iroshnikov (1963) and Kraichnan (1965). They found
that the mean magnetic field provided by large scales will, very
much unlike hydrodynamics, crucially define the dynamics on small scales.
Although this first model was isotropic, it was
eventually realized that the dynamics will result in anisotropy
(Montgomery \& Turner 1981, Shebalin et al 1983, Higdon 1984). This
resulted in Goldreich \& Sridhar 1995 (henceforth GS95)
model which postulated so-called ``critical balance'', i.e., maximum
anisotropy which is allowed under strong interaction. At the same
time, the concept of the dominant perpendicular cascade, that was used in
GS95, has been validated in an analytical perturbative theory of
so-called weak Alfv\'enic turbulence (Galtier et al. 2000,
2002)\footnote{Although a successful analytical theory, the weak
  Alfv\'enic turbulence can rarely be applied to real-life MHD
  turbulence as it would require weak, relatively isotropic
  fluctuations on outer scale and a strong mean field. Even when these
  conditions are satisfied, the strength of the interaction, which can
  be estimated as $\delta v_l L/v_A l$, where $L/l$ is the anisotropy
  of the perturbation, will increase down the cascade and break the
  applicability of the theory. In real life ISM turbulence the
  perturbations of the magnetic field are of the order of the mean
  field which makes the turbulence strong from outer scale
  inwards. For phenomenological treatment of weak Alfv\'enic
  turbulence, see also Lazarian \& Vishniac 1999.}.

While hydrodynamic turbulence have only one energy cascade, the
incompressible MHD turbulence has two, due to the exact conservation of
the Elsasser (oppositely going wave packets') ``energies''. This can
be also formulated as the conservation of total energy and
cross-helicity\footnote{The latter, $\int {\bf v}\cdot {\bf B}\, d^3x $ is a quantity conserved in the absence of dissipation.}.  The situation of zero total cross-helicity has been
called ``balanced'' turbulence as the amount of oppositely moving
wavepackets balance each other, the alternative being ``imbalanced''
turbulence.  Most of the above studies concentrated on the balanced
case, and, without exception, the GS95 model, which is the strong
cascading model with critical balance, can only be kept
self-consistent assuming balanced case.
The real life turbulence, however, is almost always imbalanced,
such as in situations when one has a strong localized source of
perturbations (the Sun for solar wind or central engine for AGN jets),
but also due to inhomogeneity of energy sources for turbulence
(supernovas and stellar winds in the ISM) and the tendency of the
decaying turbulence to become increasingly more imbalanced with
time. Moreover, the purely balanced MHD Alfv\'enic turbulence
can not be understood as it is, without understanding of the more general
imbalanced case. This is due to the fact that turbulence is a
stochastic phenomena with all quantities fluctuating and every piece
of turbulence at any given time can have imbalance in it. In this
respect, while the mean-field Kolmogorov model can be expanded to
include intermittency, the mean field GS95 model can not.

Imbalanced turbulence, or ``turbulence with non-zero cross-helicity''
has been discussed long ago by a number of authors
(Dobrovolny et al. 1980, Matthaeus \& Montgomery 1980, 
Grappin et al. 1983, Pouquet et al. 1988, Biskamp 2003 and refs. therein).
These papers testified that the non-zero cross-helicity modifies the turbulence.
Although these studies correctly reproduced separate cascades
for energy and cross-helicity, they were based on then-popular
models of MHD turbulence and later it became evident that these models are problematic.
For example, the closure theory of isotropic
turbulence (Pouquet, Frisch \& Leorat 1976) which reproduced IK model can be
rejected by both theory and numerics\footnote{The artificial term for ``relaxation of triple correlations'', that was necessary to uphold local isotropy in this model, happen
to be {\it larger} than real physical nonlinear interaction.
Also numerics show that MHD turbulence is locally anisotropic (Cho \& Vishniac 2000,
Maron \& Goldreich 2001, Cho, Lazarian \& Vishniac 2002)}.
Another class of models were based on so-called two-dimensional MHD
turbulence that, as we now know, is unable to reproduce important
properties of the real three-dimensional turbulence, such as critical
balance.

Recently several models for the strong imbalanced turbulence have been
proposed (e.g., Lithwick, Goldreich \& Sridhar 2007, Beresnyak \&
Lazarian 2008, Chandran 2008).  As long as the full self-contained
analytical model for strong turbulence continues to elude discovery, direct
numerical simulations (DNS) will be an inspiration and guidance to
theorists. While the Reynolds numbers in those simulations are fairly
modest (800-4000 are the best to-date), some of the robust quantities
can be measured and used as a guidance to reject theories.
While in this paper we are fully aware of these limitations,
we present the most robust statistical measures, such
as total energies, dissipation rates and second-order
structure function (or its equivalent, the spectrum)
and the anisotropy derived from it. We
leave the study of higher order statistics (and intermittency) to the
future studies. Our paper is written on the premise that one cannot
fully {\it confirm} a model using three-dimensional DNS due to their
fairly modest resolution, but from these DNS one can collect enough
numerical evidence to {\it reject} a model.

This paper expands parameter space of the preliminary
numerical results reported in Beresnyak \& Lazarian (2008a).
A short introduction into theories
of imbalanced turbulence is in \S 2. Numerical code,
the simulation setup, and the establishment of the stationary state
are explained in \S 3. The discussion of the structure function
calculated parallel to the magnetic field (parallel SF), a quantity
which is a key for understanding statistical anisotropy of MHD turbulence
is given in \S 4. In this section we explain various methods
to define the local guiding field and the influence to the measurement
of the parallel SF. Scale-dependent anisotropies derived from structure functions
as well as power spectra are described in \S 5.
Polarization alignment briefly discussed in \S 6.
Final comparison with models as well as observational data are in \S 7.
We summarize our findings in \S 8.

\section{Theoretical considerations}

The original GS95 model was based on the renormalization rule for
Alfv\'en wave's frequency called ``critical balance''. The necessity of
such renormalization can be seen from a rigorous theory of weak
Alfv\'enic turbulence (Galtier et al 2000, 2002) that predicts so-called
``perpendicular cascade'', i.e. the result that nonlinear interaction
of Alfv\'enic waves conserve these wave's frequencies and only
transverse structure of the wave packet is affected. As
perpendicular cascade proceeds to small scales, the applicability of
weak interaction breaks down, and Alfv\'enic turbulence becomes
strong. In this situation GS95
assumed that the frequency of the wavepacket can not be smaller than
the inverse lifetime of the wavepacket, estimated from nonlinear
interaction. In their closure model GS95 have an explicit ad-hoc term
that allows for the increase of the wave frequency. The
scale-dependency of this term is based on the assumption of turbulence
locality (i.e. there is one characteristic amplitude of perturbation
pertaining to each scale).  In the imbalanced case, however, we have
two such characteristic amplitudes and the choice for frequency
renormalization becomes unclear (GS95)\footnote{We assume that
imbalanced turbulence is ``strong'' as long as the applicability
of weak Alfv\'enic turbulence breaks down. This requires that
at least one component is perturbed strongly. In the imbalanced
turbulence the amplitude of the dominant component is larger, so
that in the transition to strong regime the applicability
of weak cascading of the subdominant component breaks down first.}.
Any theory of
strong imbalanced turbulence, which is qualified for serious
consideration, must deal with this problem. Let us first demonstrate
that a direct generalization of GS95 for imbalanced case does not
work, namely if we assume that the frequency renormalization for one
wavepacket is determined by the amplitude of the oppositely moving
wavepacket. Indeed, in this situation the wave with small amplitude
(say, $w^-$) may only weakly perturb large amplitude wave $w^+$ and
the frequency of cascaded $w^+$ will conserve. On the other hand,
$w^+$ may strongly perturb $w^-$ and $w^-$'s frequency will be
determined as $w^+_l/l$\footnote{Throughout this paper we assume that $w^+$ is
the larger-amplitude wave. This choice, however, is purely arbitrary and corresponds
to the choice of positive versus negative total cross-helicity.}.
This creates an inconsistency of the
local cascade where both wavepackets must have comparable parallel and
perpendicular scales. In order to deal with this fundamental
inconsistency, a new physical assumption must be adopted. Due to this
fact, in the paper we mostly discuss three models of strong imbalanced
turbulence, Lithwick, Goldreich \& Sridhar (2007) (LGS07), Beresnyak
\& Lazarian (2008a) (BL08a), Chandran (2008) (C08), that clearly
state: a) the problem above; b) the new physical assumptions being
adopted; c) an internally consistent physical model that follows from
these assumptions and d) the full predictions of the turbulence spectra
and anisotropy. In the discussion section of this paper we also
mention other models that claim to have predictions on the strong
imbalanced turbulence. We test these incomplete models by comparison
with our numerics whenever possible.

\def\L{{\Lambda}}
\def\l{{\lambda}}

\subsection{LGS07 and C08 models}

LGS07 resolve the inconsistency explained above by assuming that the strong
wave $w^+$ is also cascaded strongly and its frequency is simply equal to
the frequency of the weak wave, i.e. the critical balance for strong wave
uses the amplitude of the strong wave itself ($w^+\L=v_A\l$).
In other words, the anisotropies of the waves are identical.
The formulae for energy cascading are strong cascading formulae, i.e.

$$ \epsilon^\mp=\frac{(w^\mp(\l))^2 w^\pm(\l)}{\l}.$$

This lead to the prediction $w^+/w^-=\epsilon^+/\epsilon^-$.
However, this prediction, together with assumption
of the locality of cascading, lead to a contradiction on
viscous scale\footnote{LGS07 does not discuss transition to viscous scales.}
where the nonlinear cascading rates must smoothly transit
into viscous dissipation rates. This require $w^+=w^-$ on
the dissipation scale, so called ``pinning'' (Lithwick \& Goldreich 2003).

C08 is a complicated quantitative theory of advection-diffusion
cascading that have several rules to determine the diffusion of the 
waves in the parallel direction, which are analogous to the frequency
renormalization of GS95. In effect, in the case of strong turbulence,
the C08 rules lead to equal or very close anisotropies for both waves.
Unlike LGS07, however, C08 does not have a strong cascading for
both waves, but (again, in effect) it has:

$$ \epsilon^\mp=A^\mp\frac{(w^-(\l))^2 w^+(\l)}{\l},$$

where the coefficients $A^\pm$ depend on the spectral slopes of $w^\mp$.
Although the theory of C08 do not explicitly assume local cascading,
in effect it produces such a locality as long as the ratio of
dissipation rates (energy fluxes) $\epsilon^+/\epsilon^-$ is not very large
(the critical value is around two), therefore C08 also requires
pinning on the viscous scale.

\begin{figure}
\plotone{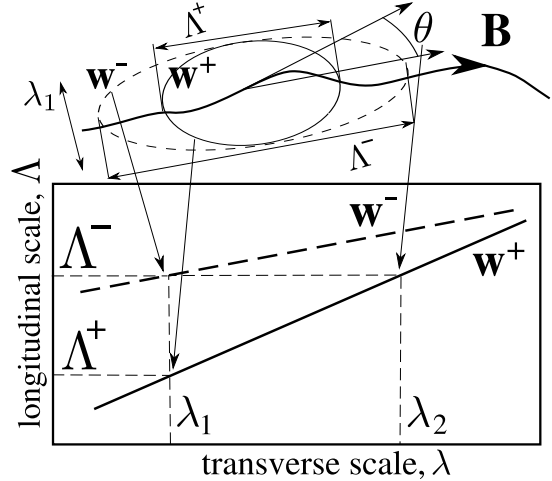}
\caption{Upper: a ${\bf w}^+$ wavepacket, produced
by cascading by ${\bf w}^-$ wavepacket is aligned with respect
to ${\bf w}^-$ wavepacket, but misaligned with respect
to the local mean field on scale $\lambda_1$, by the angle $\theta$.
Lower: the longitudinal scale $\L$ of the wavepackets,
as a function of their transverse scale, $\l$; $\L^+$, $\L^-$, $\l_1$, $\l_2$
are the notations used in this paper. Modified from BL08a.}
\label{anis_cartoon}
\end{figure}

\subsection{BL08 model}

BL08a relaxes the assumption of local cascading for the strong component $w^+$,
while saying the $w^-$ is cascaded in a GS95-like way. In BL08a picture
the waves have different anisotropies (see Fig. \ref{anis_cartoon}) and the $w^+$ wave
actually have smaller anisotropy than $w^-$, which is opposite to what
a naive application of critical balance would predict.
The anisotropies of the waves are determined by
$$w^+(\l_1)\L^-(\l_1)=v_A\l_1,\eqno{(1)}$$
$$w^+(\l_2)\L^+(\l^*)=v_A\l_1,\eqno{(2)}$$
where $\l^*=\sqrt{\l_1\l_2}$, and the energy cascading
is determined by weak cascading of the dominant wave
and strong cascading of the subdominant wave:

$$\epsilon^+=\frac{(w^+(\l_2))^2 w^-(\l_1)}{\l_1}\cdot\frac{w^-(\l_1) \L^-(\l_1)}{v_A\l_1}\cdot f(\l_1/\l_2), \eqno{(3)}$$

$$\epsilon^-=\frac{(w^-(\l_1))^2 w^+(\l_1)}{\l_1}. \eqno{(4)}$$

One of the interesting properties of BL08a model is that, unlike LGS07
and C08, it does not produce self-similar (power-law) solutions when
turbulence is driven with the same anisotropy for $w^+$ and $w^-$
on the outer scale. BL08a, however, claim that, on sufficiently
small scales, the initial non-power-law solution will transit into
asymptotic power law solution that has
$\Lambda^-_0/\Lambda^+_0=\epsilon^+/\epsilon^-$
and $\lambda_2/\lambda_1=(\epsilon^+/\epsilon^-)^{3/2}$.
The range of scales for the transition region was not specified
by BL08a, but it was assumed that larger imbalance
will require larger transition region.

\section{Numerical setup}
Incompressible MHD equations with dissipation and driving are,

\begin{equation}
\partial_t{\bf w^\pm}+\hat S ({\bf w^\mp}\cdot\nabla){\bf w^\pm}=-\nu_n(-\nabla^2)^n{\bf w^\pm}+ {\bf f^\pm},
\end{equation}

where $n$ is an order of hyperdiffusion, and ${\bf f^\pm}$ is the driving force, whose
rms values in arbitrary units are presented in Table 1.
These equations were solved by the code which is similar to one in Cho \& Vishniac (2001).
The differences include introduction of Elsasser driving, the handling of arbitrary
physical sizes of the box, regardless of numerical resolution (i.e. elongated boxes for
sub-Alfv\'enic turbulence) and significant improvements in numerical efficiency.
Our code is pseudospectral, i.e., it solves ODE in time with finite difference
for each spacial Fourier harmonic, the nonlinear term being calculated in real space,
while solenoidal projection and dissipation terms were applied in Fourier space.
Pseudospectral methods in fluid dynamics has been known since 80-s (Canuto et al, 1988).
The strong point of pseudospectral codes is that they allow precise control
over dissipation and exact incompressibility.
Not only does it relieve worries with respect to grid effects and numerical
dissipation, but it also makes possible the use of hyperviscosity and hyperdiffusivity which
extends useful inertial interval although at the expense of increased (compared
to physical viscosity) bottleneck effect and a different form of the spectral slopes
(for more discussion on spectral slopes, see Beresnyak \& Lazarian 2008b (BL08b)).
The simple version of our pseudospectral code uses periodic boundary conditions.
This necessitates a discussion on whether this introduces artificial effects.
In the subsequent sections we show that a) our numerical boxes have enough parallel
size to allow for eddies of the largest parallel size, dictated by dynamics,
to exist; b) at any given time our box contains large number of independent
turbulent realizations ($\sim40$) c) the dynamical time of the eddy is several
times smaller than it takes the eddy to cross the box boundaries.

\begin{figure}
\plottwo{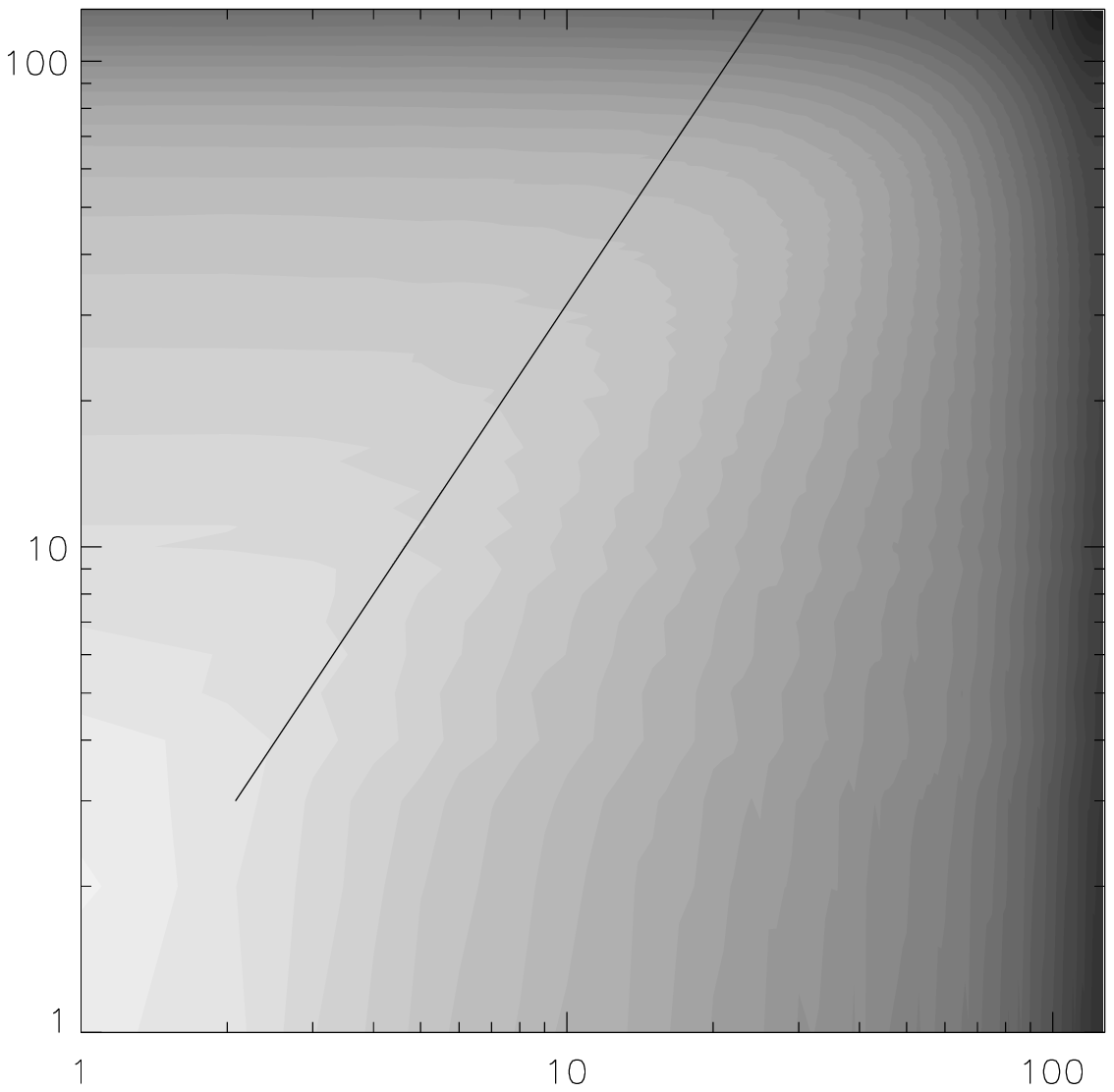}{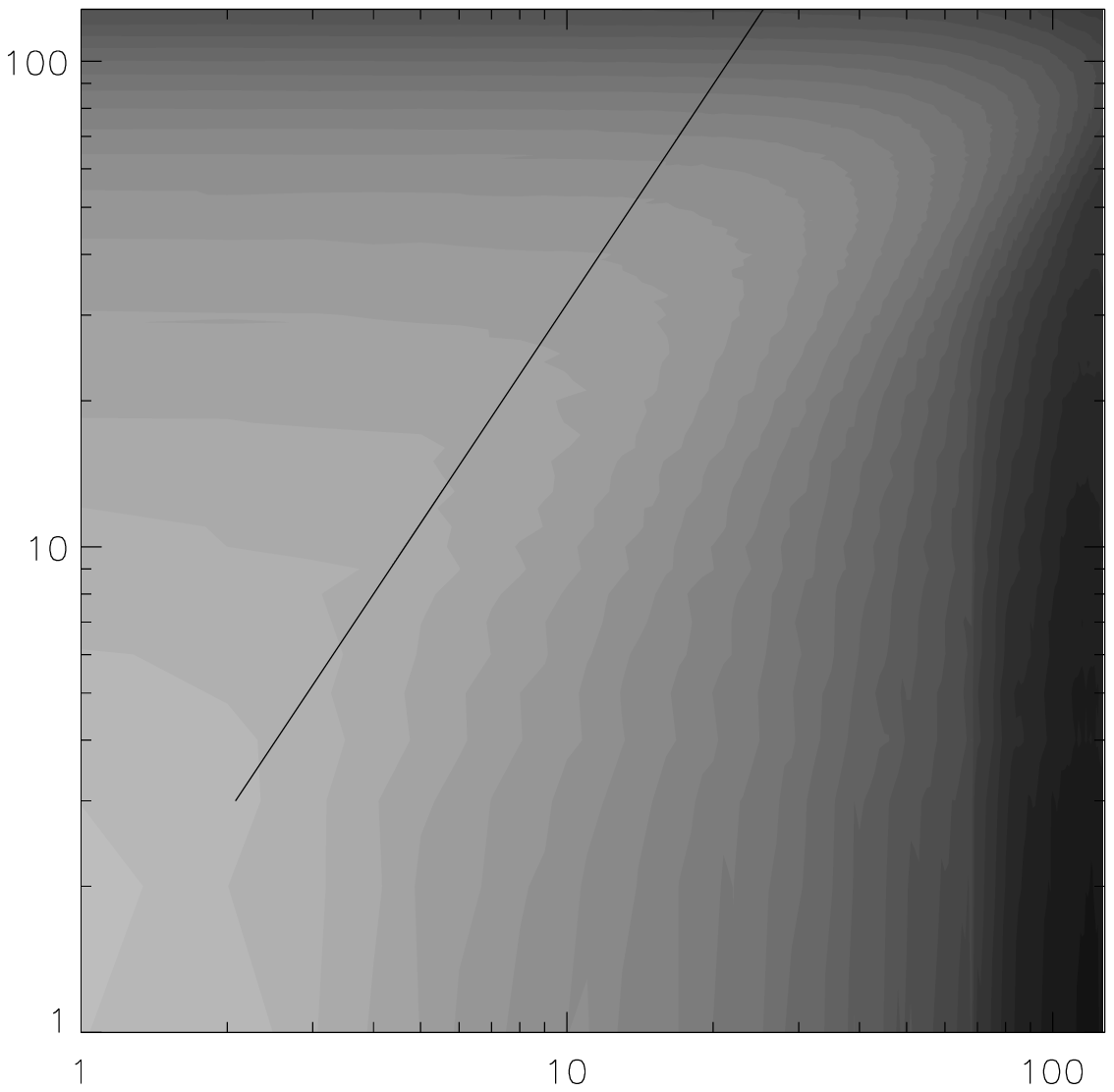}
\caption{Two-dimensional energy spectrum in sub-Alfv\'enic and trans-Alfv\'enic case.
For sub-Alfv\'enic case the abscissa $k_x$ is scaled with $1/M_A$ factor. The dashed line
indicates GS95 $k_\|\sim k_\perp^{2/3}$ anisotropy.}
\label{2dspec}
\end{figure}

\subsection{Choice of physical dimensions, numerical resolution and driving}

The sub-Alfv\'enic turbulence, where the perturbation strengths $w^\pm$ are smaller that
$v_A$ is either weak or strong but anisotropic. The critical anisotropy is determined
by the breakdown of the applicability of weak MHD turbulence which happens when
$k_\|v_A/k_\perp\delta v \sim 1$. In most of our simulations we drive turbulence on
outer scale with the same anisotropy for both wave species, so this breakdown is determined
by $\delta v$ of the strong wave. We drive turbulence on outer scale in such a manner
that the strong interaction establishes on the outer scale. Also, we provided driving
for $k=2..3.5$ which means that the maximum eddy size is several times smaller
that the box. This is to ensure that the first turbulent scales $k\approx 4$ have
more than enough space in parallel direction in case that we did not estimate
the transition into strong interaction regime correctly and the parallel scale
of the cascaded eddies is longer than we expected. The results from \S 5, however,
suggest that our choice was correct, with the largest coherent eddy size being
around 1/4 of the box size in both parallel and perpendicular directions.
We used fully predetermined stochastic driving in both Elsasser variables
\footnote{Elsasser driving is a preferred way to study intertial range of subAlfvenic
turbulence as it simulates the supply of Elsasser energies from larger eddies
of a realistic turbulence. It is important to remember that kinetic and magnetic
energies are not separately conserved by MHD equations. So when one has a pure
velocity driving in a simulation with mean field (as in Cho \& Vishniac 2001),
he will generate approximately as much magnetic perturbations due to the Alfven effect,
the result being two Alfven or pseudo-Alfven waves propagating in opposite directions.
These waves, however, would have an artificial correlation (reflected by the fact that
at $t=0$ ${\bf b=0}$). In order to use all degrees of freedom and have better
stochasticity one has to drive ${\bf w}^+$ and ${\bf w}^-$ independently.
The mechanisms by which the outer scales of a realistic, say, ISM turbulence
are driven are briefly discussed in \S 7.3 and \S 7.4.}
with a certain amplitude of the force ($f^\pm$, see Table 1), so
the energy input wasn't strictly controlled by the forcing, but rather
was calculated during the simulation\footnote{Some simulations of {\it hydrodynamic} turbulence
used negative viscosity on large scales to drive turbulence. In MHD this somewhat unphysical
approach does not work because in this case even in the balanced simulations
imbalance spontaneously occur and continue to increase without limit.}.
In addition, we developed
a driving which ensures constant energy input for both components, these
tests confirm properties of the imbalanced turbulence that were obtained
with fully stochastic driving. We used the latter for most of our simulations.

In studying sub-Alfv\'enic turbulence we adopted the approach to increase
$v_A$ by increasing $B_0$ and increase
the parallel physical size of the box $L$ by the same factor $1/M_A$ without changing the
equilibrium value of $\delta v$, so that strong interaction timescale $\l/\delta v$
stays constant and similarly the eddy transverse time $\L/v_A$ also stays constant.
Alternatively, one can keep $B_0$ constant, but decrease $\delta v$, but in this
case the timescales of sub- and trans-Alfv\'enic turbulence will be different. Also
it is harder to tune the equilibrium $\delta v$, rather than $B_0$ and $L$.

Note, that one can naively assume that due to GS95 anisotropy one needs lower numerical
resolution in the parallel direction, approximately by the ratio of the anisotropies
on the driving scale and on the dissipation scale, which is $(k_{\perp diss}/k_{\perp driv})^{1/3}$
in the GS95 model, and can be a number between 2 and 4 in a high resolution MHD simulation.
For instance, Bigot, Galtier \& Politano 2008 used 512x512x64 numerical resolution.
On the second thought this approach is not evident, since the highest
values of $k_\|$ in the {\it global reference frame} will be determined by field wandering
on the outer scale. In other words the anisotropy in the global frame will
be approximately scale-independent and the ratios of $k_{\perp diss}/k_{\perp driv}$ and
$k_{\| diss}/k_{\| driv}$ will be almost equal, which necessitates the use of NxNxN numerical
resolution, i.e. cubes, for both elongated ($M_A<1$) and cubic ($M_A\approx 1$) physical
boxes.

We confirmed this by plotting the parallel and perpendicular spectra in the global frame
and saw that the parallel spectrum protrude to almost as far as $k_{\perp max}/M_A$.
Fig. \ref{2dspec} shows how energy is distributed on the two-dimensional $k_\|,k_\perp$ plane (global
reference frame). We see that while most of the energy is in GS95 cone,
there is also plenty of energy outside of it, especially in the upper right
corner which correspond to maximum space frequencies in both parallel and
perpendicular direction. If one decide to significantly cut numerical resolution
in parallel direction he/she would incorrectly describe the dynamics on small scales.
In only one of our simulations, A1 (see Table 1) we were able to cut parallel resolution
by a moderate factor of 1.5 without sacrificing small parallel scales,
due to the relative lack of energy in parallel direction in
this particular {\it balanced sub-Alfv\'enic} case.
In all other simulations such a reduction
was not possible because most of the k-space was filled with energy.
We note that M\"uller \& Grappin (2005), by using 1024x1024x256 resolution
in their balanced sub-Alfv\'enic simulations have reduced parallel resolution
by a factor of 4, which is, most likely, too large.

For all simulations A1-A8 we used hyperviscosity and hyperdiffusivity of 6th
order ($k^6$). This choice was necessitated by the nature of imbalanced
turbulence which has shorter inertial range for dominant wave due to fairly
large cascading timescale of this wave (see \S 2). With currently
available numerical resolutions one cannot see an inertial interval of the
strong wave in a simulation with large imbalance and real ($k^2$) diffusivity.
Unfortunately, due to the bottleneck effect, hyperdiffusion have affected
spectral slopes, although the effect on anisotropy was much less. We refer to BL08b
for a comparison of turbulent simulations with normal and hyperviscosity.
Due to hyperviscosity the dissipation scale was fairly small,
the dissipation cutoff was around $k=200$ (with Nyquist frequency of 384)
for balanced simulations and about the same for weak component
in imbalanced simulations. The strong component for the most imbalanced simulations
A7 and A8 had a cutoff around $k=100$ (Fig.~\ref{spectra}).
Due to hyperviscosity
we can not uniquely define a Reynolds number of our simulations, however
viscous simulations with $Re=Re_m\approx 6000$ could provide turbulence inertial
ranges that are similar to ours.

Once we have chosen the geometry of our simulation and figured out the extend
of the perturbations on the spectral plane, the choice of timestep becomes evident.
On one hand, for the dissipation term we use integration technique
(Cho \& Vishniac 2000, see also Maron \& Goldreich 2001), and since we don't worry
too much about the precision of the dissipation term, it doesn't
limit the timestep. On the other hand, the general nonlinear term, containing
both $B_0$ and $\delta v$ can be seen as the sum of linear advection term with velocity
$v_A$ and nonlinear advection with $\delta v$, $\delta b$, etc.  
In turbulence, that is driven to be strong on the outer scale, these terms will be
of the same order if we refer to the outer scale, i.e. the terms will be
$v_A\delta v k_{\perp driv}$ and $\delta v\delta v k_{\| driv}$. On the dissipation scale
these terms will be determined by $v_A\delta v k_{\perp diss}$
and $\delta v\delta v k_{\| diss}$, which are, by the argument in the above paragraph,
again on the same order. So we can just use linear advection behavior to estimate
the timestep. This behavior in k-space is, essentially, a rotation of the phase
of the wave, in a manner of $exp(ik_\|v_At)$. In order to reproduce this rotation
numerically we need $k_{\| max}v_A\delta t$ to be smaller
than unity, such as around $0.1$, so that the code stays stable, since we
don't need good precision beyond the dissipation scale where there is no energy. 

The average dissipation rates $\epsilon^\pm$ reported in Table 1 were calculated
using a sum of the work done to the Elsasser fields, i.e. we summed
$({\bf w}^\pm+{\bf f}^\pm dt)\cdot {\bf f}^\pm dt$ at every timestep. As our code
(its nonlinear part) was energy conserving, we assume that the same amount of energy
was, on average, lost to the dissipation term.
We also confirmed these values of $\epsilon^\pm$
by using third order Chandrasekhar-Politano-Pouquet structure functions
(see, e.g., Biskamp 2003), which quantify nonlinear energy transfer.

\begin{table*}
  \large
  \begin{center}
  \begin{tabular}{c  |  c |  c |  c |  c |  c |  c |  c |  c |  c |  c |  c  | c }
    \hline\hline
Run  & $n_x\cdot n_y\cdot n_z$ & x:y:z  & $B_0$ & $f^+$ & $f^-$ & $\Delta t_0$ & $\Delta t$ & $\Delta t_1$ & Ncubes & $f^+/f^-$ & $\epsilon^+/\epsilon^-$ & $(w^+)^2/(w^-)^2$ \\
   \hline
   A1 & $512\cdot768^2$ &  10:1:1 & 10 & 0.4   & 0.4   & 186 & 10 & 4.0  & 3 & 1 & 1    &  1        \\
 
   A2 & $768^3$         &  1:1:1  &  1 & 0.44  & 0.44  & 186 & 10 & 4.0  & 3 & 1 & 1    &  1        \\

   A3  &  $768^3$      &  10:1:1 & 10 & 0.34  & 0.255 & 500 & 26 &  8.0  & 6 & 1.33 & 2.0  &  $5.5\pm 1.0$  \\

   A4  &  $768^3$      &  1:1:1  &  1 & 0.4   & 0.3   & 180 & 22 &  8.0  & 5 &1.33 & 1.7  &  $3.9\pm 0.3$  \\

   A5 &  $768^3$       &  10:1:1 & 10 & 0.16  & 0.08  & 154 & 33 & 12.0 & 6 & 2 & 7.4    &  $145\pm 10$  \\

   A6  &  $768^3$     &  1:1:1  &  1 & 0.19  & 0.095 & 40  & 33 & 12.0 &  5 &  2 & 5.4    &  $90\pm 10$  \\

   A7  &  $768^3$     &  10:1:1 & 10 & 0.06  & 0.02  & 204 & 63 & 21.0 &  8 & 3 & 16   &  $1150\pm 100$  \\

   A8 &   $768^3$     &  1:1:1  &  1 & 0.075 & 0.025 & 160 & 63 & 21.0 &  8 & 3 & 12   &  $1100\pm 100$  \\
   \hline

  \end{tabular}
  \end{center}
  \caption{Simulations of strong sub-Alfv\'enic ($B_0=10$)
           and trans-Alfv\'enic ($B_0=1$) turbulent flows.
           A1 and A2 are balanced simulations for comparison with the rest.
           A3 and A4 are slightly imbalanced, A5 and A6 are strongly imbalanced and
           A7 and A8 are very strongly imbalanced. $\Delta t_0$ is the duration of prior low
           resolution run in code units, $\Delta t$ is the duration of the high resolution run,
           interval $\Delta t_1$ in the end of high-resolution run was used for data analysis.}
  \label{experiments}
\end{table*}

\begin{figure}
\plotone{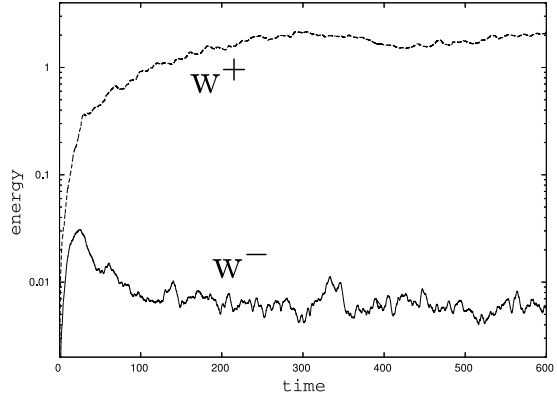}
\caption{The energy evolution in time for both Alfv\'enic modes for
$\epsilon^+/\epsilon^-=12$.}
\label{stat_en}
\end{figure}

\subsection{Establishment of the stationary state}

One of the goals of this paper is to demonstrate that a stationary state
exists for imbalanced turbulence with rather high degree of imbalance.
Note, that the local model of weak Alfv\'enic turbulence work
for imbalances of no more than $\epsilon^+/\epsilon^-=2$
(Galtier 2000, Lithwick \& Goldreich 2003), and the model of strong
imbalance turbulence of C08 also require similar limitation.

\begin{figure}
\plotone{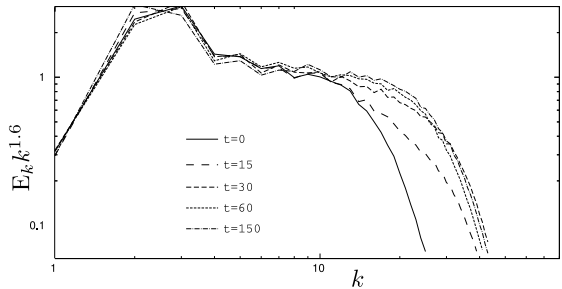}
\caption{Spectrum relaxation towards stationary state. The resolution of
the experiment was increased from $128^3$ to $256^3$. Only the spectrum of the
dominant wave is shown. The lines correspond to $t=15,30,60,150$.}
\label{stat_spec}
\end{figure}

The highest imbalance we attempted in our simulation of $\epsilon^+/\epsilon^-=16$
was essentially limited by the long times of establishment of the stationary state. 
Note that according to BL08a the dominant wave is cascaded
weakly and its cascading times could be very large. Fig. \ref{stat_en} shows the total energy
evolution for both modes for the $\epsilon^+/\epsilon^-=16$ case\footnote{Time was
measured in Alfv\'enic units, but the size of the box was $2\pi$, thus $2\pi$ was the time
for an eddy to cross the box. The time for the largest turbulent eddy to cross
itself, and also the largest eddy dynamical time ($L/v$) was around
unity, because the size of the largest eddy was a fraction of around $0.2$
or $0.3$ of the cube size (see Figs. \ref{sf2d_bal} and \ref{sf2d_imbal}).}.
The full relaxation towards
stationary state required around $300$ Alfv\'en times or $50$ crossing times.

As high imbalanced simulations proved to be so computationally expensive,
we made a second experiment, which was to take the initial state that was already
stationary and to increase the numerical resolution, which allow the spectrum to extend
to larger wavenumbers. Note, that our forcing, although stochastic, was predetermined
for each particular simulation and did not depend on numerical resolution.
Now the question was how fast the spectra will relax to
their stationary states. It turned out that the spectrum of the sub-dominant
wave relaxed almost instantly, in a one dynamical time, which is consistent
with BL08a, while for the dominant wave the relaxation time
was long. Note, the the dynamic (kinetic) timescale $l/v$ for this region
of $k$-space for the strong wave was rather small, around $0.3$. The relaxation
process is shown on Fig. \ref{stat_spec}. It took $\delta t\approx 60$ to get reasonably close
to the stationary state. We considered this experiment a success and used
the technique of increasing resolution to save computational time.

We also studied long-term evolution of nearly-balanced case when we allowed
the low-resolution version of A3 to evolve for $500$ time units. We didn't
notice any long-term trends either in three-dimensional spectrum or
in any other quantities during this run.

\section{Parallel structure function}

\begin{figure}
\plotone{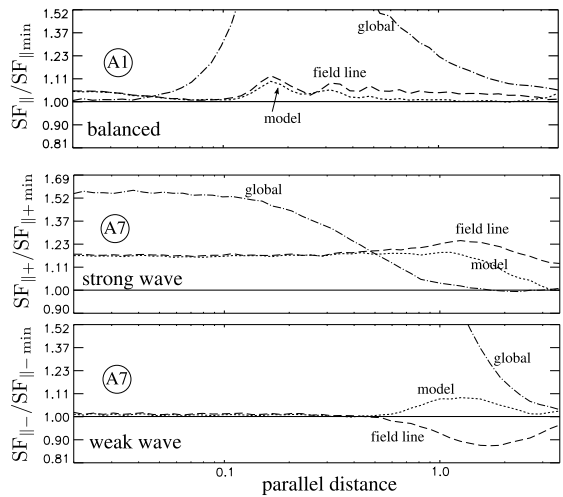}
\caption{The ratio between parallel SFs calculated with different
definition of the local field. Upper plot is for balanced simulation,
while two lower plots are for strongly imbalanced simulation.
The reference SF is ``minimum'' one described in the text. The dotted line
is ``model-dependent'' SF, dashed line is for ``following a field line'' method
and the dotted-dashed line is for constant global mean field.}
\label{sfpar_cmp}
\end{figure}

Unlike perpendicular structure function which is largely insensitive to the direction
of the local field, the definition of the local field strongly affect the parallel structure
function, which, in turn, determines the shape of the turbulent eddy. Since the latter
is the major object of study in this paper we feel that the proper explanation of this point
is due.

The simplest way to define parallel structure function, $SF_\|$, is to take samples along
the {\it global mean field.} This definition is, however, fairly bad, as it does not
take into account field wandering. We expect $SF_\|$, defined in such way, along with
perpendicular structure function to reflect the anisotropy in the {\it global frame},
which, by the effects of field wandering, as we argued in \S 2, will be similar to
the outer scale anisotropy. Therefore, such definition will effectively erase {\it
scale-dependent anisotropy} which is the property of GS95-type models.

Another way is to define local magnetic field by averaging over some scale $\l$. In this
way the parallel structure function become a function of two scales, such as

$$
SF^2_\|(w^\pm,\L,\l)=\langle(w^\pm({\bf r}-\L{\bf b}_\l/b_\l)-w^\pm({\bf r})  )^2\rangle_{\bf r},
$$

where ${\bf b}_\l$ is the magnetic field averaged over scale $\l$. We use Gaussian averaging
defined as ${\bf b}_\l=1/\l\sqrt{2\pi}\int {\bf b}({\bf r-R})\exp(-R^2/2\l^2)\, d{\bf R}$.
In order to reduce such a SF to a function of only $\L$ one can introduce a dependence
between $\L$ and $\l$, and plug in the $\l=f(\L)$ in the above equation. This
definition of $SF_\|$ will be a {\it model-dependent} though.

$$
SF^2_{\|model}(w^\pm,\L)=SF^2_\|(w^\pm,\L,f(\L))
$$

For the balanced turbulence the anisotropy was measured to be close to the one
predicted by GS95, i.e. $\L\sim\l^{2/3}$. One can, therefore, introduce a {\it reasonable}
model-dependent $SF_\|$ as taking $f(x)=const\cdot x^{3/2}$, where the constant depend
on the outer scale of the simulation. As we show below, this definition is almost perfect
for balanced turbulence, but the question is whether it does equally well for the imbalanced case.

Let us consider some {\it model-independent} ways to determine $SF_\|$. Apparently, the
first definition using global field is model-independent, but fairly bad, it correspond
to taking averaging $\lambda=\infty$. One can also take $\lambda=0$, i.e. always use local
field without any averaging. An interesting model-independent method was used in
Maron \& Goldreich 2001, where two points were chosen to lie on the same magnetic field
line. The distance $\L$ was also calculated along the line.

\begin{figure}
\plotone{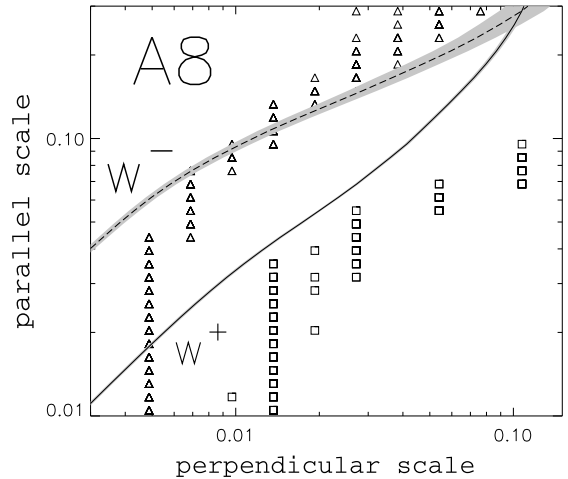}
\caption{
This plot shows values of $\l_{\rm avr}$ at which the minimum of the parallel
structure function is reached. Triangles show perpendicular scales $\l_{\rm avr}$
at which the minimum of $SF^2_\|(w^-,\L)$ is reached, while squares show perpendicular
averaging scales at which the minimum of $SF^2_\|(w^+,\L)$ is reached. 
Solid and dotted lines indicate $w^+$ and $w^-$ eddies' anisotropy
which are defined in \S 5 and presented on Fig. \ref{anis_imb}.}
\label{eddies}
\end{figure}

When we look for anisotropy
we normally want to obtain lower values of $SF_\|$. According to the eddy ansatz,
outlined in \S 2, we receive {\it lower} values of
$|{\bf w}^\pm({\bf r}-\L{\bf n(r)})-{\bf w}^\pm({\bf r})|$,
where ${\bf n}$ is a unit vector along the eddy. Therefore, the averaging of the field
that provides {\it minimum} values of $SF_\|$ approximates the direction of the eddy
alignment better, provided that there is a connection between the field direction
and eddy alignment (if there is no such connection, there will be no dependence on
the averaging scale $\l$).

So, another model independent way to define $SF_\|$ will be

$$
SF^2_{\|min}(w^\pm,\L)=\min_\l SF^2_\|(w^\pm,\L,\l).
$$

This definition not only provides us with the value of $SF_\|$, but,
giving $\l$ at which minimum is achieved, gives us a hint to how
eddies are aligned with respect to the magnetic field.

Fig. \ref{sfpar_cmp} shows a comparison between different methods to calculate
parallel SF. We plotted them relative to $SF^2_{\|min}$.
In the balanced case the three methods -- ``minimal'', ``following the field line''
and ``model-dependent'' work very well, while ``global field'' method doesn't
work. The latter confirms that turbulent eddies are
aligned with respect to local field, not the global field (Cho \& Vishniac 2000).
In the imbalanced case the situation is more complicated. For the weak component
all three ``good'' methods work very well, while for the strong wave
there is a systematic error for all methods, except ``minimal''.
This is due to the fact that in the imbalanced turbulence the
strong component ($w^+$) eddies are aligned with respect to
much larger scales of the magnetic field (see \S 2.2). Since most magnetic field
perturbation is provided by the strong wave, it follows that
the strong field eddies are aligned with their own field on a larger
scale. This directly confirms the prediction of BL08a model.
On the bottom panel of Fig. \ref{sfpar_cmp} the ``field line'' method gives
values that are smaller than ``minimal'' method. This is due
to the fact that in the field line method we measured the distance
along magnetic field, and the physical distance was actually shorter,
this allowed for smaller than $SF^2_{\|min}$ values on the outer scale
where the difference between straight-line distance and along-the-field-line
distance is significant.

\begin{figure}
\plotone{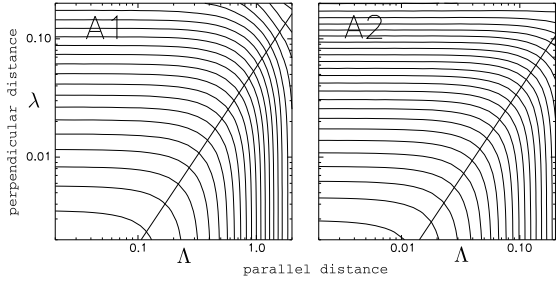}
\caption{Comparison of the SFs from trans-Alfv\'enic (left) and sub-Alfv\'enic (right)
balanced simulations. Note the difference in x axis between two plots which indicates
that A1 is approximately 10 times more anisotropic.
Contours indicate SF levels, solid line is a demonstration
of GS95 $\L\sim\l^{2/3}$ law}
\label{sf2d_bal}
\end{figure}

It turns out (Fig. \ref{eddies}) that the averaging scale
at which minimum of parallel structure is reached
for weak wave approximately corresponded to its
anisotropy, which is consistent with strong cascading hypothesis.
But for strong wave this averaging scale is {\it larger} than the
perpendicular scale dictated by anisotropy. In other words, the
eddies of the strong wave are aligned with respect to the magnetic
field which is averaged on {\it larger scale} than the eddy's
own perpendicular scale. This is consistent with the BL08a model.

\section{Spectra and anisotropies}

\begin{figure}
\plotone{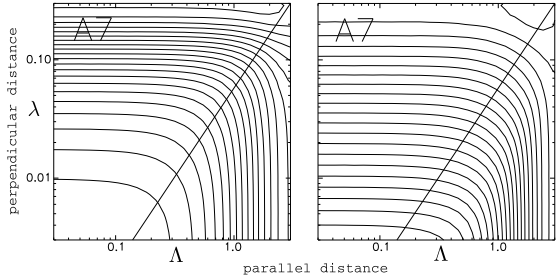}
\caption{Comparison of the SFs from dominant (left) and subdominant (right) Alfv\'enic
waves from $\epsilon^+/\epsilon^-=16$ imbalanced simulations. The anisotropies
of components are notably different.
The solid line is the GS95 $\L\sim\l^{2/3}$ law}
\label{sf2d_imbal}
\end{figure}

\begin{figure}
\includegraphics[width=\columnwidth]{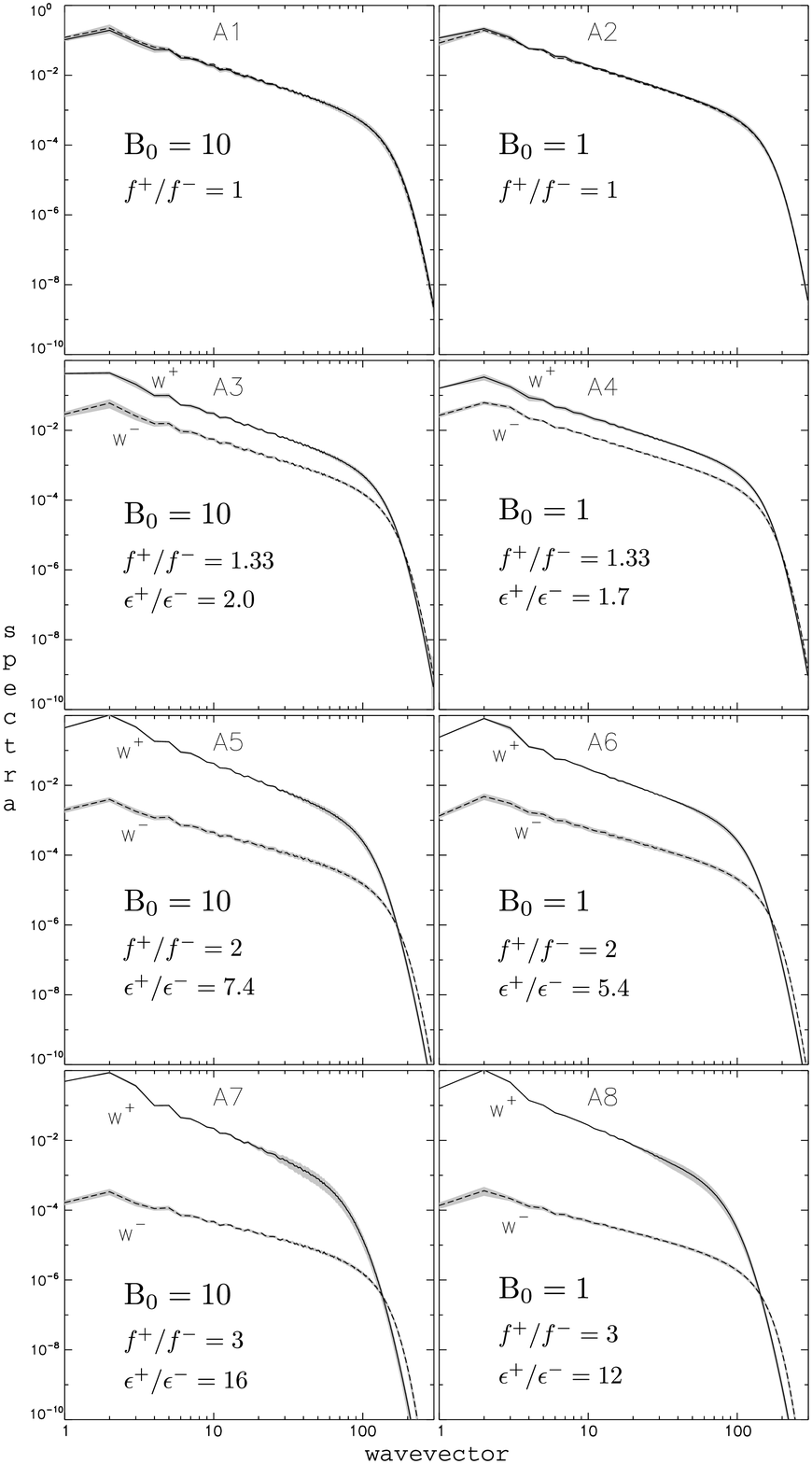}
\caption{Spectra for all data, gray shows mean-square fluctuations in time.}
\label{spectra}
\end{figure}

\begin{figure}
\plotone{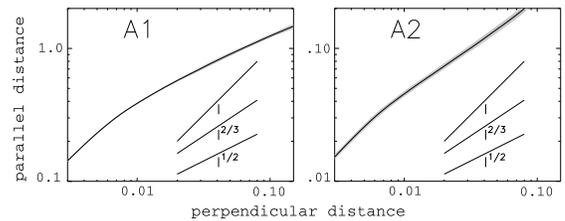}
\caption{Anisotropies
for balanced simulations.}
\label{anis}
\end{figure}

We calculated two-dimensional (depending on parallel and perpendicular distances)
second order structure functions with respect to the local field using
``model dependent'' definition of the local field from previous section. Although
this method slightly underestimates anisotropy, according to Fig. \ref{sfpar_cmp},
it works fairly well.
The structure functions were calculated using all available ``stationary state''
datacubes, i.e. were averaged over time.
The contours of these SFs for balanced simulations A1 and A2 are presented on Fig. \ref{sf2d_bal}
 and for imbalanced simulation A7 on Fig.  \ref{sf2d_imbal}. Fig.  \ref{sf2d_bal} shows SFs for total energy i.e.
it is summed over $w^+$ and $w^-$. These figures basically validate our assumptions
from \S 3 regarding physical and computational dimensions of the box. We see that,
according to expectations, trans-Alfv\'enic A2 is almost isotropic on outer scale but becomes
progressively anisotropic towards small scales, while, as we expected,
sub-Alfv\'enic A1 has approximately 10:1 anisotropy on outer scale, and increases
towards small scales. If we decrease anisotropy of A1 by a factor of 10 by rescaling
x-axis we almost reproduce A2, the difference is mostly being on the outer scale
(this difference is easier to see on Fig \ref{anis}.). Fig. \ref{sf2d_imbal} shows SFs
for two separate components $w^+$ and $w^-$ in the strongly imbalanced case of A7.
The anisotropy on outer scale is approximately 10:1 for both components,
which validates our choice of computational box. This anisotropy increases
towards small scale, but in a different fashion for each component. We see
the anisotropy of strong wave is almost 5 times smaller on dissipation scales.

Fig.~\ref{spectra} shows so-called three-dimensional angle-summed spectra for both components
in all simulations. These spectra are obtained by summation of spectra
over solid angle for all wavevectors with the same magnitude $k$. It can be related
to three-dimensional angle-averaged spectra by dividing by $k^2$. In the sub-Alfv\'enic
cases A1, A3, A5 and A7, this spectrum is almost identical to the so-called perpendicular
spectrum, which takes into account only structures perpendicular to the magnetic field
and is the main target of prediction of GS95 model. As they are almost identical
we did not have to plot perpendicular spectrum separately. Another definition of
spectrum which depend only on the magnitude of the wavevector is the so-called
one-dimensional spectrum (see, e.g. Monin \& Yaglom 1965). This spectrum is
less sensitive to the bottleneck effect. We refer to the paper of Beresnyak \& Lazarian (2008b)
(henceforth BL08b)
for a more thorough comparison between one-dimensional and 3D spectra and discussion
on bottleneck effect.

\begin{figure*}
\plotone{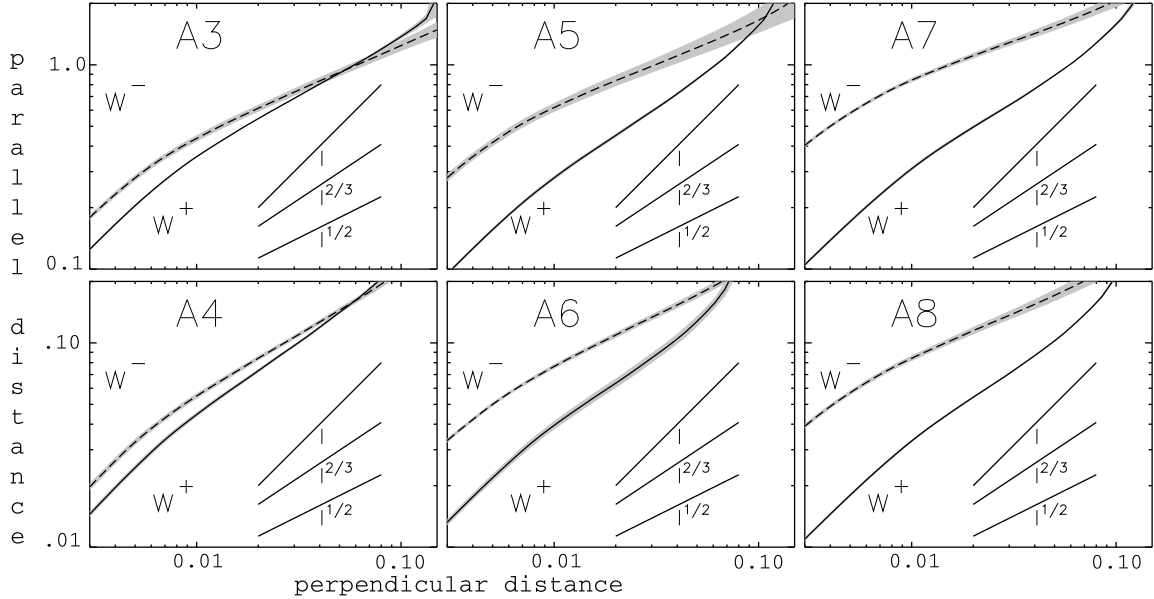}
\caption{Anisotropies
for imbalanced simulations. The mapping of $\L(\l)$ is explained in \S 5.
The difference in anisotropy between $w^+$ and $w^-$ increases with increasing
imbalance.}
\label{anis_imb}
\end{figure*}

In Fig.~\ref{spectra}, the two bottom plots have relatively large variation (gray areas)
on the end of the $w^+$ spectra. This is due to the fact that in A7 and A8
we barely reached stationary state (for more discussion, see Fig. \ref{stat_spec} and
\S 3.2) in the high-resolution run.

Fig. \ref{spectra} shows spectral slopes between -1.12 and -1.93 with balanced simulations
having slopes as flat as -1.37\footnote{The slopes for one-dimensional
spectra are steeper, with balanced slope of -1.45 and the most
imbalanced slopes -1.97 for strong wave and -1.22 for weak wave}.
The GS95 prediction is Kolmogorov's $-5/3\approx-1.67$,
while Boldyrev 2006 prediction is -1.5. The flat slopes observed in real data
are most certainly due to rather strong bottleneck effect
seen in simulations with hyperviscosity. In the imbalanced case the predictions
are following: LGS07 predicts -1.67 slopes for all 8 cases; C08 predicts
-1.67 for balanced cases A1 and A2, A5-A8 are outside of the applicability
of his model, A3 and A4 must show very different slopes -- approximately
-1 for weak component and -3 for strong component, also C08 predicts
pinning on dissipation scales i.e. spectra should converge on dissipation scale
\footnote{LGS07 does not discuss transition to viscous scales, but,
as a model of local cascading, it must have pinning on viscous scale.
The pinning, however, is impossible within the framework of LGS07, as the latter
predicts $w^+/w^-=\epsilon^+/\epsilon^-$.}. The ratios of the total energies (see Table 1) are
predicted as following: C08 -- A5-A8 are outside of applicability
of his model, A3 and A4 should have very large imbalance $(w^+)^2/(w^-)^2$
of at least a 1000, while 4-6 is actually observed; LGS07 predicted/observed --
A3: 4/5.5, A4: 3/3.9, A5: 55/145, A6: 30/90, A7: 260/1150, A8: 144/1100.
We see that in this respect deviations from LSG07 predictions are small
for small imbalances but fairly large for large imbalances.
BL08a argues that if one drives turbulence with the same anisotropy
on outer scale (as in these simulations) the anisotropies of the components
will diverge towards small scales, and this solution will not be
self-similar (and not power-law). However, BL08a makes predictions regarding
{\it local} slopes even in this case. This can be seen from (1) which
is a classic critical balance between weak wave anisotropy and strong wave
amplitude and (4) which is strong cascading of the weak wave.
We do not expect relations between slopes based on (4) to hold, because
it is strongly influenced by bottleneck effect (BL08a also predicts -1.67
slopes for balanced case). However, there is some dependence between
energy slope and anisotropy slope, similar to what BL08a predicts.
Namely, it follows from (1) that shallower anisotropy slope for $w^-$
means steeper spectral slope for $w^+$ which is observed. Also,
from (4), steeper spectral slope for $w^+$ also means shallower spectral
slope for $w^-$ which is also observed.

The anisotropy was measured in the following manner. First, the parallel
and perpendicular second order structure functions were calculated, then we
found equal values of parallel and perpendicular SFs and in this way the mapping
or function between independent variables, parallel or perpendicular scales were
created. This function is plotted on Figs \ref{anis}, \ref{anis_imb}
with shades of gray indicating RMS
fluctuations in time. This definition of $\Lambda(\lambda)$ mapping can be understood
from two-dimensional plot of SF, e.g. Fig. \ref{sf2d_bal}, when one follows a contour of SF
and finds which parallel scale correspond to particular perpendicular scale.
We see that for the imbalanced case anisotropy curves have different slopes
and diverge from outer scale where they are equal (this is dictated by driving)
to smaller scales where they are different.

We devoted \S 4 to the discussion of the measurements of the parallel
structure function which was used in the above definition of the anisotropy
curves. Although it might appear that each and every definition 
produce a different anisotropy curve, the major difference is between global and local
definition of the field direction, while all local methods (``field line'',
``model-dependent'' and ``minimal''), also dubbed ``good'' in \S 4, give very
similar results. In fact, these is no perceivable qualitative difference
between anisotropy curves obtained by either local methods. This could be explaned
by Fig. 4 middle and bottom panels where the quantitative differences
between methods are small, but on the other hand, the dependence of $SF_\|$ on scale
is strong ($\sim l_\|$). Also, in the middle panel, the difference is mostly by a constant,
which will only give a slight shift of the anisotropy plot.
All in all, the claim that anisotopy curves will diverge by a factor of 3 to 4 in
strongly imbalanced simulations stay true regardless of the ``local'' method used.
We rejected ``global field'' method as it does not reveal scale dependent anisotropy
-- a groundbase of GS95 model. It is worth noting that LGS07, C08 and BL08 use
GS95 as a basis and smoothly transit to GS95 in the balanced limit.
There is a wealth of theoretical arguments why the SFs has to be measured
with respect to the local field (Cho \& Vishniac 2000, Maron \& Goldreich 2001, etc).
We also would like to note that aside from driven simulations described in this
paper we also observed a significant difference in $w^+$ and $w^-$ anisotropies
in a {\it decaying} imbalanced simulations.  

C08 and LGS07 both predict identical GS95 anisotropy for
both modes, which is inconsistent with simulations.
BL08a predicts diverging anisotropy, most notably, with
stronger wave having smaller anisotropy, which is consistent with simulations.
The value of the differences, however, do not reach the asymptotic value
of $\epsilon^+/\epsilon^-$ which may be attributed to the short inertial
range.

\section{Polarization Alignment}
Aside from energy-type statistics (second order SFs and spectra)
one can measure so-called alignment effects which are, in a sense
deviations from the assumptions of independent randomness
of fluctuations included in mean-field models.
These were discussed in Boldyrev (2005)
and numerically discovered for the first
time in Beresnyak \& Lazarian (2006).

\begin{figure}
\plotone{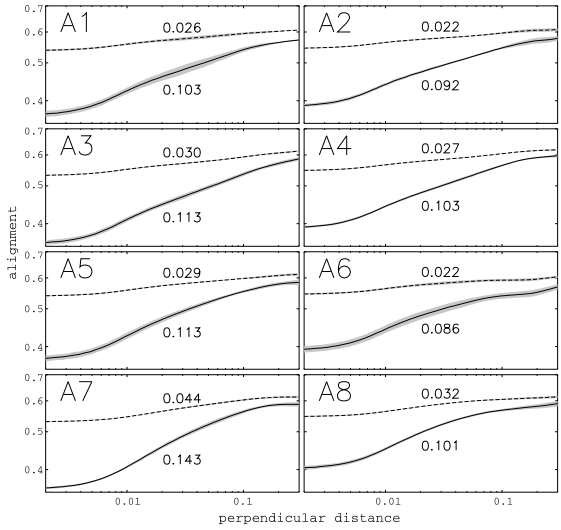}
\caption{Polarization angle alignment (dashed) and polarization intermittency (see definition
in the text). Numbers are power-law slopes determined in the middle of the inertial interval.}
\label{alignment}
\end{figure}

Fig \ref{alignment}. shows two measures of alignment -- the angle polarization alignment
$AA=\langle|\sin\theta|\rangle$ (dashed line), where
$\theta$ is an angle between Elsasser variables perturbations
$\delta {\bf w}^+=\delta {\bf v}+\delta {\bf b}$ and
$\delta {\bf w}^-=\delta {\bf v}-\delta {\bf b}$
and ``polarization intermittency'' $PI=\langle | \delta w^+ \delta w^- \sin \theta
|\rangle /\langle |\delta w^+ \delta w^-|\rangle$ (solid line).

For the more detailed numerical study of alignment effects
we refer to Beresnyak \& Lazarian (2006) and BL08b.
The potential effect of alignment on the energy
cascade was discussed in Boldyrev (2005, 2006),
although this effect has not yet been convincingly confirmed by
numerics (see BL08b for a discussion and refs).

\section{Discussion}
\subsection{Comparison with models}
We briefly mentioned the tentative nature of {\it confirming a model} in \S 1,
we also claimed the ability of numerics to {\it reject} {\it some models}
on the basis of robust quantities. Although a theory can make a wide variety
of predictions, only few of those can be effectively attacked by numerics.
One of the quantities that is notoriously hard to measure in DNS is the
spectral slope of turbulence. A difference between $-3/2$ slope and $-5/3$
can be masked by a variety of effects such as bottleneck effect, driving,
and so on. In contrast, the quantities such as Kolmogorov constant are
fairly easy to obtain and quickly converge with increasing resolution.
In fact, modest resolutions such as $128^3$ give reasonably precise
estimates of this constant. This is due to the fact that the {\it total
energy} and the {\it total dissipation rate} are easy to measure, to get
a statistical average, and also are free of uncertainties of interpretation.
What sort of models can be judged
on the basis of these quantities? Such are the models of {\it local cascading}
where the cascade rate depend only on the characteristic quantities of,
say, $w^\pm_l$ on a particular scale $l$ (and, possibly, weakly depend on the
spectral slope). In this case numerics only has to reproduce a one or two
steps of such cascading to obtain reasonable dissipation rate based
on a particular total energy. In this sense our testing does fairly well,
as we mostly consider models of local cascading (LGS07, C08,
Perez \& Boldyrev (2009), Podesta \& Bhattacharjee (2009)), at the same
time, with $768^3$ numerical resolution we reproduce five to six binary
steps in $k$-space.

\begin{figure}
\plotone{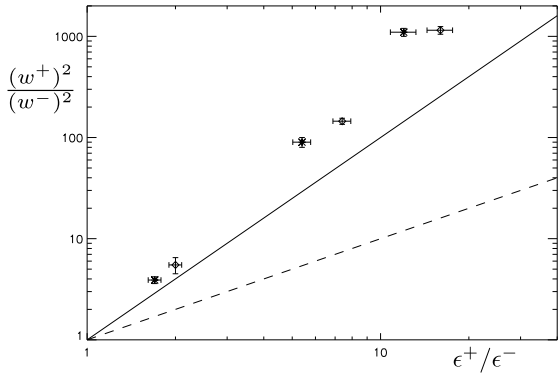}
\caption{Total energy ratio versus dissipation rate ratio (see also Table 1).
Diamonds: subAlv\'enic simulations, stars: transAlv\'enic simulations,
solid line: LGS07 prediction, dashed line: Perez \& Boldyrev (2009) prediction.
There is no simple formula for C08 model, but the energy ratio is expected to
become very large when dissipation rate ratio approaches the critical value
of around two.}
\label{en_vs_eps}
\end{figure}

Our numerical data strongly contradict to three models
of imbalanced turbulence, namely LGS07, C08, and Perez \& Boldyrev (2009).
In particular, two of these models C08, and Perez \& Boldyrev (2009) show
gross inconsistencies between observed and predicted energy ratios vs dissipation
ratios. Indeed, C08 must have a huge energy ratio (of around a 1000)
in simulations with $\epsilon^+/\epsilon^-$ close to two (A3 and A4),
while a modest ratios of 4 and 6 are observed. Perez \& Boldyrev (2009)
does extremely bad in cases with large imbalances. A7 and A8 has an
energy ratios of around a 1000, while predicted quantities are 16 and 12.
LGS07 does a much better job on energy ratios, but still fails the
A7 and A8 (large imbalance) tests, see Fig.~\ref{en_vs_eps}.

Furthermore, LGS07 and C08 have predictions regarding eddie anisotropies.
Both of these models predict equal anisotropies for $w^+$ and $w^-$
while a different anisotropies are observed.
Aside from these inconsistencies, we also note that C08 predicts pinning
at the dissipation scale, which is not observed. C08 also predicts
that the strong wave has to have steeper spectral slope than
the weak wave. This corresponds to numerics qualitatively, but
not quantitatively, indeed, according to C08, A3 and A4 must
have a huge slope difference of around 2, while the real difference is
around 0.12.

Although it is harder to confirm a model rather than to reject
a model by direct numerical simulations, we see that there is a qualitative
agreement between BL08a and numerics. Most of the features
predicted by BL08a are observed in simulations, namely a) the anisotropies
of the waves are different and strong wave anisotropy is smaller; b) while
the weak wave eddies are aligned with respect to the local field on the same
scale as the eddy, the strong wave eddies are aligned with respect to
a larger-scale field, (Fig. \ref{eddies});
c) the energy imbalance is higher than in the case
when both waves are cascaded strongly (Table 1), which suggest that
the strong wave is cascaded weakly; d) the dissipation scales for
the weak and the strong waves are different, namely the inertial range
for weak wave is longer (Fig. \ref{spectra}) which is what predicted
by BL08a; e) there is no ``pinning'' at the dissipation scale which
suggest nonlocal cascading.

We note that there is no quantitative agreement between the difference
in anisotropies in the ``asymptotic power-law solutions'' of BL08a
and simulations (c.f. \S 2 and Fig. \ref{anis_imb}). This is probably
due to the fact that asymptotic power-law solutions have not been
established within our inertial range.

Aside from three models with detailed strong imbalanced theories
that we described above, LGS07, BL08a and C08, recently two papers
also made predictions regarding strong imbalanced turbulence.
Perez \& Boldyrev (2009) predicted that $(w^+/w^-)^2=\epsilon^+/\epsilon^-$
which comes to much stronger contradiction with A7 and A8 than LGS07
and as such should be discarded. We note parenthetically that measurements
of the stationary levels of energies for $w^+$ and $w^-$ are
the most robust and model independent of all other measurements.
Another paper, by Podesta \& Bhattacharjee (2009) is a modification
of Perez \& Boldyrev (2009) which tries to resolve a huge discrepancy
of the latter. Unfortunately, their theory has an arbitrary parameter
that can be tuned to change the prediction for $(w^+/w^-)^2$ and,
therefore, it can not be tested numerically.
We also note that two aforementioned papers do not have clearly stated predictions
for anisotropy, which also limits our ability to test them. 

\subsection{Comparison with earlier simulations}
Maron \& Goldreich (2001) and Cho et al. (2002) performed 3D numerical
simulations of decaying imbalanced turbulence. In these simulations
the perturbations obtained as a result of balanced driven MHD modeling
were separated into oppositely moving flows of Elsasser variables and
one of the flows was arbitrary decreased in amplitude.
The authors observed the increase of the damping time for
the strong component. They also observed the increase of turbulence imbalance as
the turbulence was evolving. Naturally, no stationary state was
achievable for the imbalanced turbulence induced this way.

%
%

\subsection{Role of homogeneous turbulence driving}

Properties of imbalanced turbulence may depend on how it is driven. All three major
models of imbalanced turbulence that we discuss above (LGS07, BL08a and C08)
describe imbalanced turbulence driven homogeneously
through the volume. The properties of such turbulence
may differ if the sources of driving are localized.
To illustrate this fact, in the Appendix~A we discuss
a toy model of {\it weak} imbalanced turbulence driven
inhomogeneously at boundaries. Note that this approach is very different
that the DNS approach of the rest of this paper.
Being the model of weak turbulence,
this toy model describes only a {\it perpendicular cascade} just like its precedessor
Lithwick \& Goldreich (2003), which considered homogeneous case.
However, the stationary states of these inhomogeneous and homogeneous cases
substantially differ.
We also expect to see substantial differences for the homogeneous and inhomogeneous driving
for the {\it strong} imbalanced turbulence, but since such a study
with the use of the DNS will be complicated and computationally expensive,
we defer this discussion to future publications. 

\subsection{Effects of compressibility}

The simulations presented here are of incompressible MHD
turbulence. Not only full compressible MHD equations have
more degrees of freedom (incompressible approach exclude fast mode
perturbations), but compressibility substantially changes the properties of
turbulence when sonic Mach numbers are of around unity or higher.
However, numerical studies in Cho \& Lazarian (2002, 2003) showed
that coupling of Alfv\'enic and magnetosonic waves in strong turbulence
is not as strong as was expected, which can be due to the fast
non-linear decay of Alfv\'enic eddies.

In the imbalanced turbulence the strong wave is long-lived. Therefore, one
can expect the imbalanced turbulence to be more affected by coupling
of incompressible and compressible motions. Density
inhomogeneities present in the compressible fluid can act as mirrors
reflecting waves and decreasing the degree of turbulence imbalance.
Parametric instabilities (see, e.g., Del Zanna et al. 2001) can develop
in the compressible fluid, decreasing the imbalance. Thus, stationary
states with high degree of imbalance may not be attainable in
compressible fluids. Further research in this direction is
necessary.

\section{Summary}

In the paper above we have performed MHD numerical simulations of the homogeneously driven imbalanced turbulence and have shown that:\\
1. Stationary states exist for rather high degree of imbalance. \\
2. For large imbalances, the ratio of the amplitudes disagrees with the predictions in LGS07.\\
3. Rough correspondence of the expectations and measurements are obtained for the model in BL08a,
but more studies and testing is necessary.

\section{Appendix A: Inhomogeneous Weak Imbalanced Turbulence -- A Toy Problem.}
Although we understand that the models of imbalanced turbulence
are in their making and even the case of homogeneous imbalance turbulence
is still being investigated, we decided to add this appendix that probes
into a different problem, namely, inhomogeneous Alfv\'enic turbulence.

This is motivated by the fact that imbalanced turbulence often
appears in inhomogeneous setting, i.e. when one has a strong
localized source of waves, such as the Sun in solar wind turbulence.
In this respect the theory of homogeneous imbalance turbulence
should be well-applicable to small scales where the time scales
are much smaller than outer timescales. However, on larger scales,
homogeneity could be broken. In this section we probe the
situation when it is broken. This can be used
as a guidance to what extent the observations of the solar wind
can be considered as restrictive measurements with respect to the
theories of imbalanced turbulence.

\begin{figure}
\plotone{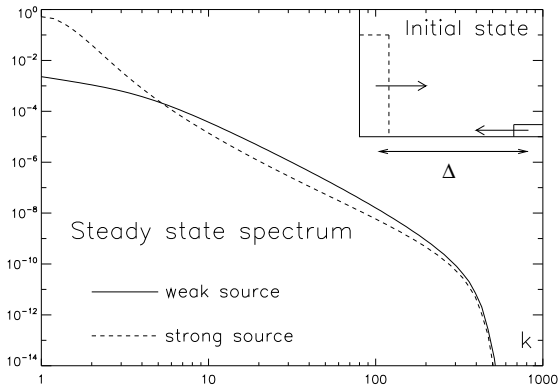}
  \caption{Spectrum of up (solid) and downgoing (dashed) waves,
in the middle between the wave sources.}
\label{inhom}
\end{figure}

Previous studies of weak Alfv\'enic turbulence
(e.g. Galtier et al 2000, Lithwick \& Goldreich 2003)
considered spatially homogeneous case, which presume
that $w^+$ and $w^-$ are driven homogeneously in space.
This assumption is often violated in nature, specifically
when we have a strong directional source of perturbations
such as the Sun, which drives one component (let us say, $w^+$)
and the other component is being generated at a certain distance
from the source by means of, e.g., reflection from
density inhomogeneities. This problem is closely related
to the problem of the imbalance, however, in most of this
paper we considered idealized spatially homogeneous case.
Dissipation of waves in turbulence (Beresnyak \& Lazarian 2008c)
is also related to this problem.

We realize, that by relaxing homogeneity we
considerably broaden the physical scope of the model, to
the point that we might be unable to draw clear conclusions
on the nature of inhomogeneous turbulence. In this situation
we preferred to take a first step by considering a toy problem
of weak turbulence with two wave sources separated
by a certain distance.

One remarkable new property that could arise from such formulation
is that the cascading {\it might not reach the dissipation scale}.
This could happen if, e.g., the sources of the waves are too
close to each other. One of the questions that we are asking
in this appendix, is whether it is possible that only one wave
component, e.g. $w^-$ is dissipated but the other is simply distorted.

We used simplified equations of weak cascading, the diffusive
one dimensional k-space equations for weak perpendicular cascade
from Lithwick \& Goldreich (2003) which
were expanded to one spacial dimension by introducing
advection in space, making them an advection-diffusion equations
with advection happened in real space, while diffusion represented
energy cascading in k-space. We used open boundaries, so that the
non-cascaded waves were allowed to freely escape through them.

By solving those equation numerically, changing the distance
between wave sources we found that the approximate equality
of the energies of the waves at the smallest scale which is
reached by cascading is rather robust feature, that is produced
by cascading itself. Note that Lithwick \& Goldreich 2003 assumed that
this ``pinning'' is due to the dissipation term.
In our toy model, however, regardless of whether an actual dissipation
took place in the system, the spectra were ``pinned'' on small scales.

Another possibility that was opened by the inhomogeneous formulation
was to drive wave sources with arbitrary power and still obtain
a stationary state. In the homogeneous formulation the stationary state
was not possible if the rate of energy driving $\epsilon^+/\epsilon^-$
was larger than 2 (Lithwick \& Goldreich 2003). Our inhomogeneous toy model
can deal with larger imbalances since the waves were allowed to escape
through the open boundaries of the box.
Fig. \ref{inhom} shows spectrum for both types of waves in such case
of strong imbalance, and incomplete cascading mentioned above
(energy hasn't reached dissipation scale). Note, that unlike
DNS with periodic boundaries from the main body of this paper,
where energy can only be lost through dissipation,
in this case energy could be lost due to open boundaries
and the stationary state could be achieved without any physical
dissipation actually taking place.

Fig.~\ref{inhom} demonstrates a feature which was not observed
in Lithwick \& Goldreich (2003), namely that imbalance
reverses on scales an order of magnitude smaller than
the driving scale. This unexpected feature
is fairly robust in the case ``incomplete''
cascading and strong imbalance. This suggest that inhomogeneous
imbalanced turbulence could be much more complicated that its
homogeneous counterpart and further research is necessary.

\acknowledgments
AB thanks IceCube project for support of his research. AB thanks Teragrid project
for providing computational resources.
AL acknowledges the  NSF grant ATM-0648699, AST-0808118 and the support from
the Center for Magnetic Self-Organization in Laboratory and Astrophysical
Plasma.

\end{document}